\documentclass[default,iicol]{sn-jnl}

\jyear{2022}%

\theoremstyle{thmstyleone}%
\newtheorem{theorem}{Theorem}

\theoremstyle{thmstyletwo}%
\newtheorem{example}{Example}%

\theoremstyle{thmstylethree}%
\newtheorem{definition}{Definition}%

\begin{document}

\title[The Quantum-House Effect]{The Quantum-House Effect: Filling the Gap Between Classicality and Quantum Discord}

\author*{\fnm{Tamás} \sur{Varga}}\email{tvarga@q-edu-lab.com}

\affil{\orgname{Papafut Quantum}, \orgaddress{\state{Zurich}, \country{Switzerland}}}

\abstract{We introduce the quantum-house effect, a non-local phenomenon which apparently does not require quantum discord to be present. It suffices for the effect if neither subsystem of a bipartite system is in a pure state. This way, the quantum-house effect completely fills the gap between trivial correlations and quantum discord. However, we discuss why the situation is more subtle than that, by showing that in a concrete cryptographic setting called "the quantum-house game", the ability to produce quantum discord is in fact necessary for the quantum-house effect to work. Then, we suggest a principle called "quantum detachment" to characterize where quantumness in general departs from classicality, based on the information a physical system contains about itself. The quantum-house effect is demonstrated on SpinQ Gemini, a 2-qubit liquid-state NMR desktop quantum computer.}


\maketitle

\section{Introduction}\label{sec1}

Quantum entanglement was discovered nearly a century ago \cite{einstein}, and has become a signature effect of quantum mechanics. Schrödinger called entanglement \textit{the} characteristic trait of quantum mechanics, which alone embodies the difference between quantumness and classicality \cite{schrodinger}.

In the past decades, entanglement has turned out to be a key resource in quantum information processing as well \cite{nielsen,schumacher}. In particular, several authors pointed out that entangled states play an essential role in achieving exponential speed-up in certain quantum-computing algorithms \cite{ekert}, and others hinted that in the absence of entanglement one should not talk about "true" quantum computation, but rather a simulation thereof \cite{braunstein}.

It was against this backdrop that in 2001 quantum discord was discovered \cite{ollivier}. Quantum discord is a measure of the quantumness of correlations between two quantum subsystems: non-zero discord certifies the presence of non-classical correlations. As one would intuitively expect, entangled states always have non-zero discord. Surprisingly, however, there also exist bipartite states which aren't entangled but still have non-vanishing discord. So entanglement isn't necessary for the non-classicality of correlations, non-zero discord already suffices.

But can we take this even further? Are there bipartite states with zero discord that exhibit non-classical correlations? The short answer is yes, and in Section \ref{sec2} we introduce the quantum-house effect, which pushes the territory of non-classical correlations right into the realm of product states, as every bipartite product state where neither subsystem is in a pure state is capable of this non-local effect. However, we also argue that the situation is more subtle than that, and in a cryptographic sense quantum discord is still required for the quantum-house effect to work, even when bipartite states with zero discord are used.

The quantum-house effect can be considered as an extension of locally non-effective unitary operations, first proposed in \cite{fu}, and further investigated in \cite{datta}.\footnote{The present work came about independently of \cite{fu}. The core idea we arrived at is basically the same, but our original motivation was educational, to explore quantum effects that can be demonstrated on SpinQ Gemini.}

\begin{figure}[t]
\centering
\includegraphics[width=0.7\columnwidth]{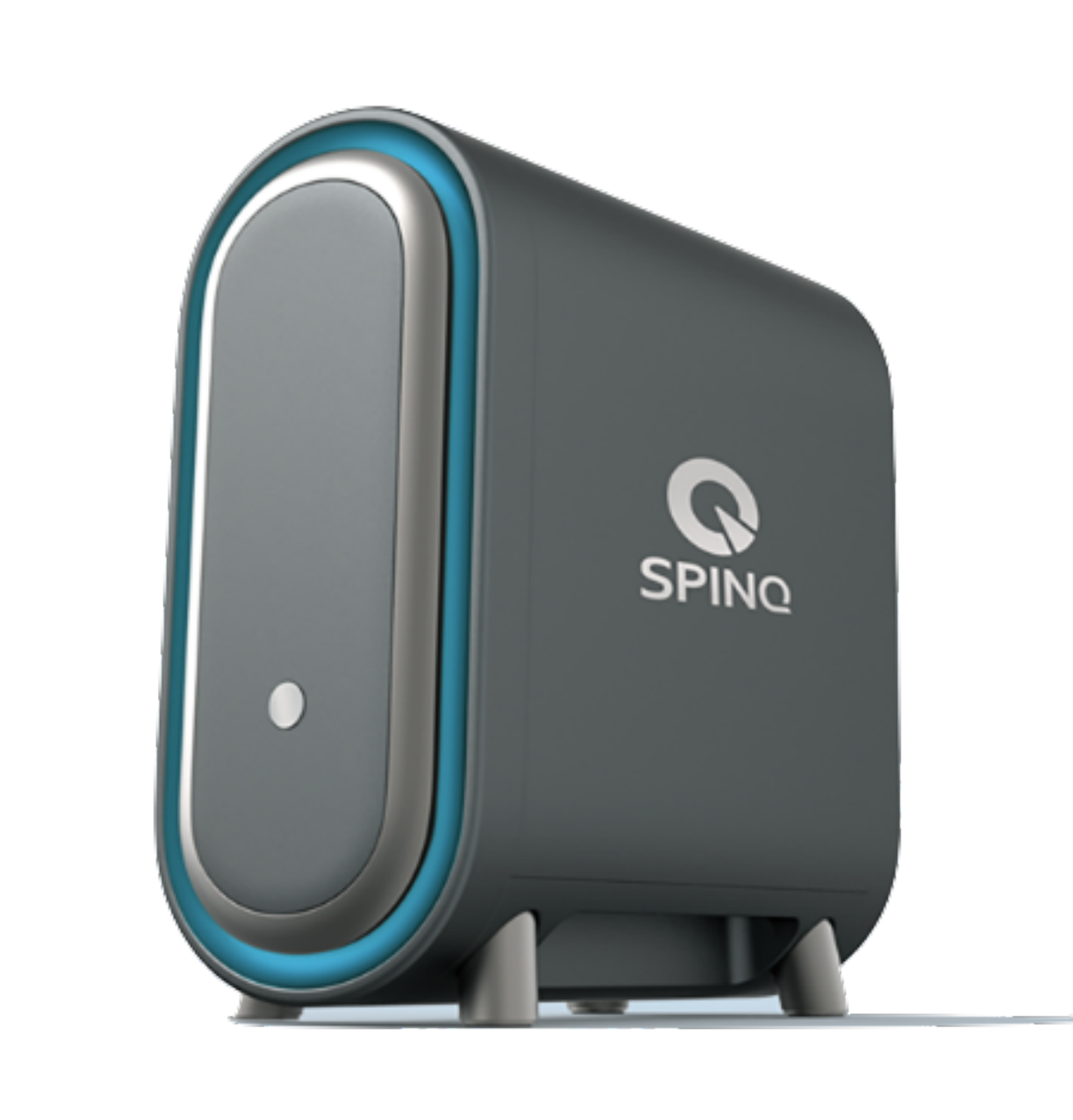}
\caption{SpinQ Gemini, a 2-qubit NMR desktop quantum computer. Source: \url{https://www.spinq.cn/}}\label{figgemini}
\end{figure}

The paper is organized as follows. Sections \ref{sec2} and \ref{sec4} present theoretical results, while in Section \ref{sec3} a demonstration of the quantum-house effect is given using SpinQ Gemini, a 2-qubit liquid-state nuclear magnetic resonance (NMR) desktop quantum computer \cite{hou}, shown in Fig. \ref{figgemini}. Then, cryptographic aspects are touched upon in Section \ref{sec5}. Finally, Section \ref{sec6} summarizes our findings and suggests a principle to differentiate quantumness from classicality in general.

\section{Theory - Part 1}\label{sec2}

Quantum discord is reviewed briefly in Subsection \ref{subsec21}, after which the quantum-house effect is introduced in Subsection \ref{subsec22}.

Throughout this paper, we work with a bipartite system $AB$, where subsystem $A$ belongs to Alice, and subsystem $B$ to Bob. The overall quantum state of $AB$ is represented by the density matrix $\rho_{AB}$, while those of $A$ and $B$ are given by the partial trace formulas $\rho_A=\mathrm{tr}_B( \rho_{AB})$ and $\rho_B=\mathrm{tr}_A( \rho_{AB})$, respectively.

\subsection{Quantum discord}\label{subsec21}

Quantum discord for $\rho_{AB}$ is a non-negative number, denoted by $\delta_{AB}$ \cite{ollivier}. In this paper, we are only interested in whether or not the discord is vanishing, which can be characterized as follows \cite{ollivier}:

\begin{theorem}[Non-vanishing discord]\label{thmnvd}
$\delta_{AB}>0$ if and only if any complete projective measurement performed on subsystem $A$ would perturb $\rho_{AB}$ for a bystander who is unaware of the measurement.
\end{theorem}

That is, if Alice secretly performs \textit{any} complete projective measurement on her subsystem $A$, then for Charlie who is unaware of that measurement, the overall state of system $AB$ will change to some $\rho'_{AB}\ne\rho_{AB}$, giving him a chance to figure out that Alice did something.

\begin{example}\label{exnvd}
Let's take the EPR pair $\rho_{AB}=\frac{1}{2}(\vert 00\rangle+\vert 11\rangle)(\langle 00\vert+\langle 11\vert)$, where Alice owns the first qubit, Bob the second. To keep things simple, let Alice measure the first qubit in the computational basis. For Charlie, unaware of the measurement, the result will be that $AB$ is either in the state $\vert 00\rangle$ or $\vert 11\rangle$, with probability $\frac{1}{2}$ each. We can thus write for Charlie $\rho'_{AB}=\frac{1}{2}\vert 00\rangle\langle 00\vert+\frac{1}{2}\vert 11\rangle\langle 11\vert$, which is different from $\rho_{AB}$.\footnote{Charlie could get $\rho'_{AB}$ to any precision by quantum-state tomography, if the experiment is repeated enough times.} So Charlie may figure out that Alice did something. Later, in Example \ref{exdis2}, we'll see that Alice could have chosen \textit{any} basis for her measurement, it would always perturb the overall 2-qubit state. That is, the EPR pair has non-zero discord.
\end{example}

On the other hand, vanishing discord $\delta_{AB}=0$ implies there \textit{is} a way for Alice to secretly measure $A$ via complete projection such that Charlie surely wouldn't notice anything. This scenario is what one would expect to be always possible in the classical world. In this sense, vanishing discord captures a notion of "classicality", while non-vanishing discord $\delta_{AB}>0$ certifies the presence of non-classical correlations that cannot exist in a classical setting.

\begin{example}\label{exvd}
Let's continue where we left off at Example \ref{exnvd}, and let this time $\rho_{AB}=\frac{1}{2}\vert 00\rangle\langle 00\vert+\frac{1}{2}\vert 11\rangle\langle 11\vert$. Now, if Alice measures the first qubit in the computational basis, it won't change the 2-qubit density matrix for Charlie who is unaware of the measurement, i.e. $\rho'_{AB}=\rho_{AB}$ will hold. So Charlie won't have chance to figure out if Alice has measured or not. That is, $\rho_{AB}$ has zero discord.
\end{example}

It's easy to see that the presence of discord can be characterized in terms of superposition, which was also noted in \cite{ollivier} and \cite{modi}:

\begin{theorem}[Discord is superposition]\label{thmdis}
$\delta_{AB}=0$ if and only if $\rho_{AB}=\sum_{i}p_i\vert a_i\rangle\langle a_i\vert\otimes\rho_B^i$, where $p_i\ge 0$, $\sum_i p_i=1$ and $\{\vert a_i\rangle\}$ is an orthonormal basis of $A$.
\end{theorem}

Thus, vanishing discord means $\rho_{AB}$ can be produced essentially without superposition in subsystem $A$, relying solely on the orthogonal states of the single basis $\{\vert a_i\rangle\}$.\footnote{One can arrive at the formula in Theorem \ref{thmdis} from a hardware perspective as well. It can be shown that the pseudo-entangled states \cite{hou} produced by SpinQ Gemini cannot be expressed by such a formula, i.e. without superposition.}

\begin{example}\label{exdis1}
Since $\frac{1}{2}\vert 00\rangle\langle 00\vert+\frac{1}{2}\vert 11\rangle\langle 11\vert=\frac{1}{2}\vert 0\rangle\langle 0\vert\otimes\vert 0\rangle\langle 0\vert+\frac{1}{2}\vert 1\rangle\langle 1\vert\otimes\vert 1\rangle\langle 1\vert$, according to Theorem \ref{thmdis} it is an example of zero discord, with $p_1=p_2=\frac{1}{2}$, $\vert a_1\rangle=\vert 0\rangle$, $\vert a_2\rangle=\vert 1\rangle$, $\rho_B^1=\vert 0\rangle\langle 0\vert$ and $\rho_B^2=\vert 1\rangle\langle 1\vert$.
\end{example}

\begin{example}\label{exdis2}
Since the EPR pair is an entangled state, it cannot be expressed by the separable-state formula in Theorem \ref{thmdis}, and thus $\delta_{AB}>0$ must hold (see also Example \ref{exnvd}).
\end{example}

Finally, if a device is able to produce just two (different) non-orthogonal states of $A$, say, $\vert u\rangle$ and $\vert v\rangle$ with $0\lt\vert\langle u\vert v\rangle\vert\lt 1$, it's already enough to create quantum discord. E.g. the state $\rho_{AB}=0.6\vert u\rangle\langle u\vert\otimes\vert 0\rangle\langle 0\vert+0.4\vert v\rangle\langle v\vert\otimes\vert 1\rangle\langle 1\vert$ has non-vanishing discord.

\subsection{The quantum-house effect}\label{subsec22}

\begin{figure}[t]
\centering
\includegraphics[width=1.0\columnwidth,trim={1.55cm 0 1.55cm 0},clip]{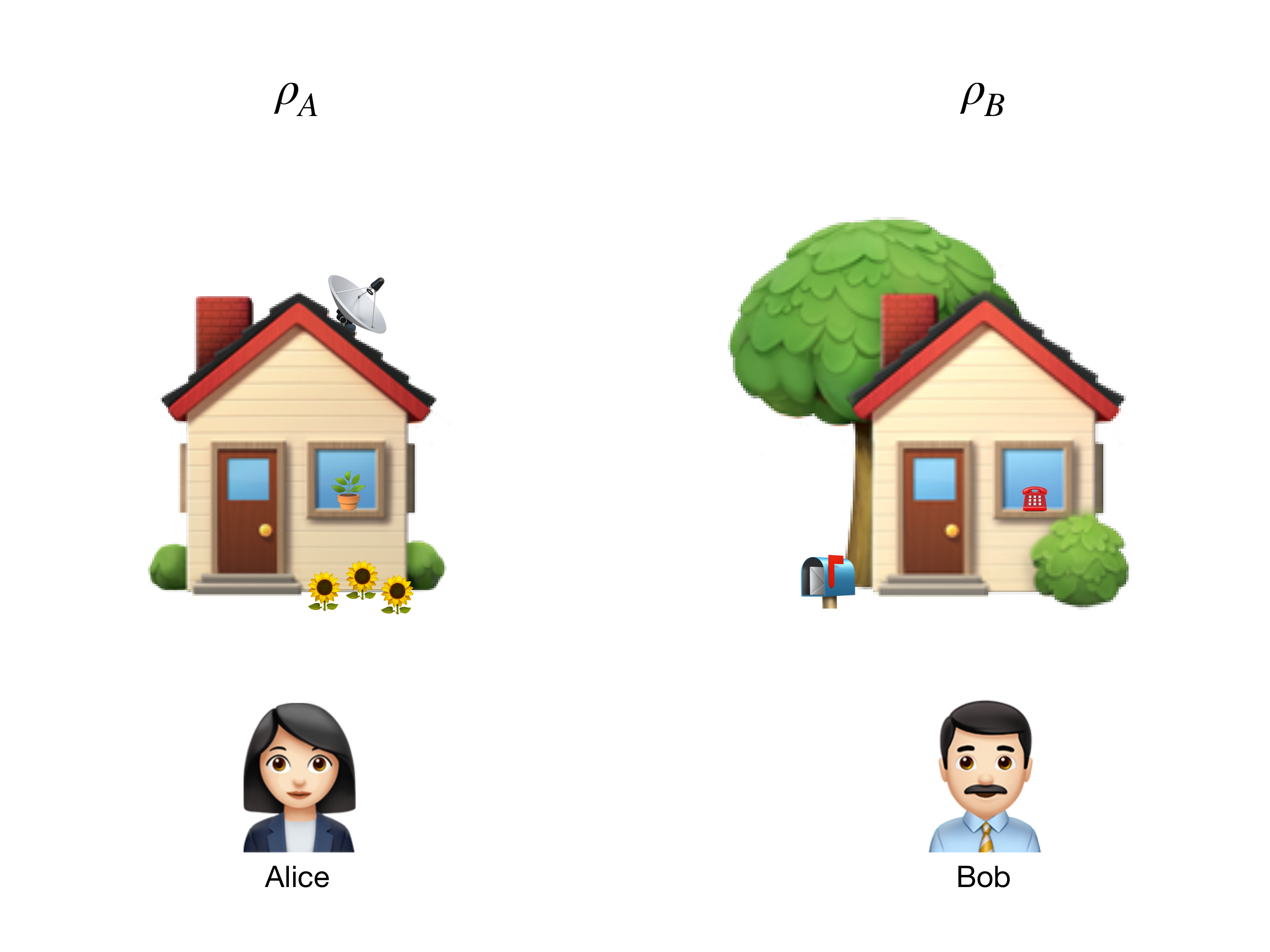}
\caption{Analogy for the quantum-house effect. Charlie enters Alice's house and either makes a change or does nothing. However, Alice cannot figure out, by examining solely her own house, whether or not Charlie has made any change. But if she joins forces with Bob, the two of them \textbf{together} may figure it out. This is because the change is such that it affects only the joint state $\rho_{AB}$ of the two houses, but not the state $\rho_{A}$ of Alice's individual house.}\label{figaqh}
\end{figure}

Let's start with the technical definition:

\begin{definition}[Quantum-house effect]\label{defqhe}
The quantum-house effect is the phenomenon when an operation on subsystem $A$ changes the overall state of system $AB$, but not that of $A$.
\end{definition}

That is, the state of $AB$ changes to some $\rho'_{AB}\ne\rho_{AB}$, while the state of $A$ remains $\rho_A$. The "operation on $A$" can be anything, as long as it's performed inside Alice's lab: measurements, unitaries, or combinations thereof, with or without ancilla system.\footnote{When only unitaries are allowed without ancilla, such an operation not affecting the state of $A$ is called a locally non-effective unitary operation, see \cite{fu} and \cite{datta} for details.} The only restriction is that during the operation we don't have access to Bob's lab where subsystem $B$ resides.

The quantum-house effect is a non-classical phenomenon, because in the classical world any such operation in Alice's lab which doesn't change the state of $A$ cannot change the state of $AB$ either. (By "state" in the classical world we mean a \textit{complete} and \textit{objective} description of a physical system in itself that can be in principle established by an experimenter who is given access to the system in his/her local lab.) Furthermore, it is also a non-local effect in the sense that the impact of the operation spans two locations, i.e. labs, although according to our classical intuition it should be confined just to Alice's lab.

\begin{example}\label{exqhe}
Imagine this time it is Charlie who secretly measures, in the computational basis, Alice's qubit of an EPR pair $\rho_{AB}=\frac{1}{2}(\vert 00\rangle+\vert 11\rangle)(\langle 00\vert+\langle 11\vert)$. Then, since Alice and Bob are unaware of Charlie's action, the new 2-qubit state for them will be $\rho'_{AB}=\frac{1}{2}\vert 00\rangle\langle 00\vert+\frac{1}{2}\vert 11\rangle\langle 11\vert$, which is different from $\rho_{AB}$. That said, the state of Alice's individual qubit stays the same: $\rho'_A=\rho_A=\frac{1}{2}\vert 0\rangle\langle 0\vert+\frac{1}{2}\vert 1\rangle\langle 1\vert$. So Alice has no chance to figure out by herself that Charlie did something. But \textit{together} with Bob, they may figure it out!
\end{example}

For a more intuitive understanding, an analogy is shown in Fig.~\ref{figaqh}. In this analogy, subsystems $A$ and $B$ are houses of Alice and Bob, respectively. Then, a change made secretly by Charlie on Alice's house may only be detected by Alice and Bob together, but not by Alice alone examining her own house.

Here is a theorem about the relationship to quantum discord:

\begin{theorem}[Discord implies quantum-house]\label{thmdiqh}
If $\delta_{AB}>0$, then the quantum-house effect can be achieved with $\rho_{AB}$.
\end{theorem}

\begin{proof}
Let $\rho_A=\sum_{i}p_i\vert a_i\rangle\langle a_i\vert$ be the spectral decomposition of $\rho_A$. Now, if Alice's subsystem $A$ gets measured in the $\{\vert a_i\rangle\}$ basis, then for a bystander unaware of the measurement the state of $A$ won't change, but due to Theorem \ref{thmnvd} the state of the overall system $AB$ will.
\end{proof}

The quantum-house effect is also possible with zero discord, as it can be seen in the example below. This way, it extends the notion of non-classicality offered by quantum discord, to a wider range of bipartite quantum systems.

\begin{example}\label{exqhezd}
Let $\rho_{AB}=\frac{1}{2}\vert 00\rangle\langle 00\vert+\frac{1}{2}\vert 11\rangle\langle 11\vert$. We saw in Example \ref{exdis1} that this state has zero discord, due to Theorem \ref{thmdis}. Now, if we apply a Pauli-$X$ gate on the first qubit, the overall 2-qubit state will change to $\rho'_{AB}=\frac{1}{2}\vert 10\rangle\langle 10\vert+\frac{1}{2}\vert 01\rangle\langle 01\vert$, which is different from $\rho_{AB}$. On the other hand, the state of the first qubit remains $\rho'_A=\rho_A=\frac{1}{2}\vert 0\rangle\langle 0\vert+\frac{1}{2}\vert 1\rangle\langle 1\vert$.
\end{example}

\section{Demonstration on SpinQ Gemini}\label{sec3}

In this section, we'll showcase the quantum-house effect on the SpinQ Gemini 2-qubit NMR desktop quantum computer \cite{hou}.

\begin{figure}[t]
\centering
\includegraphics[width=0.95\columnwidth]{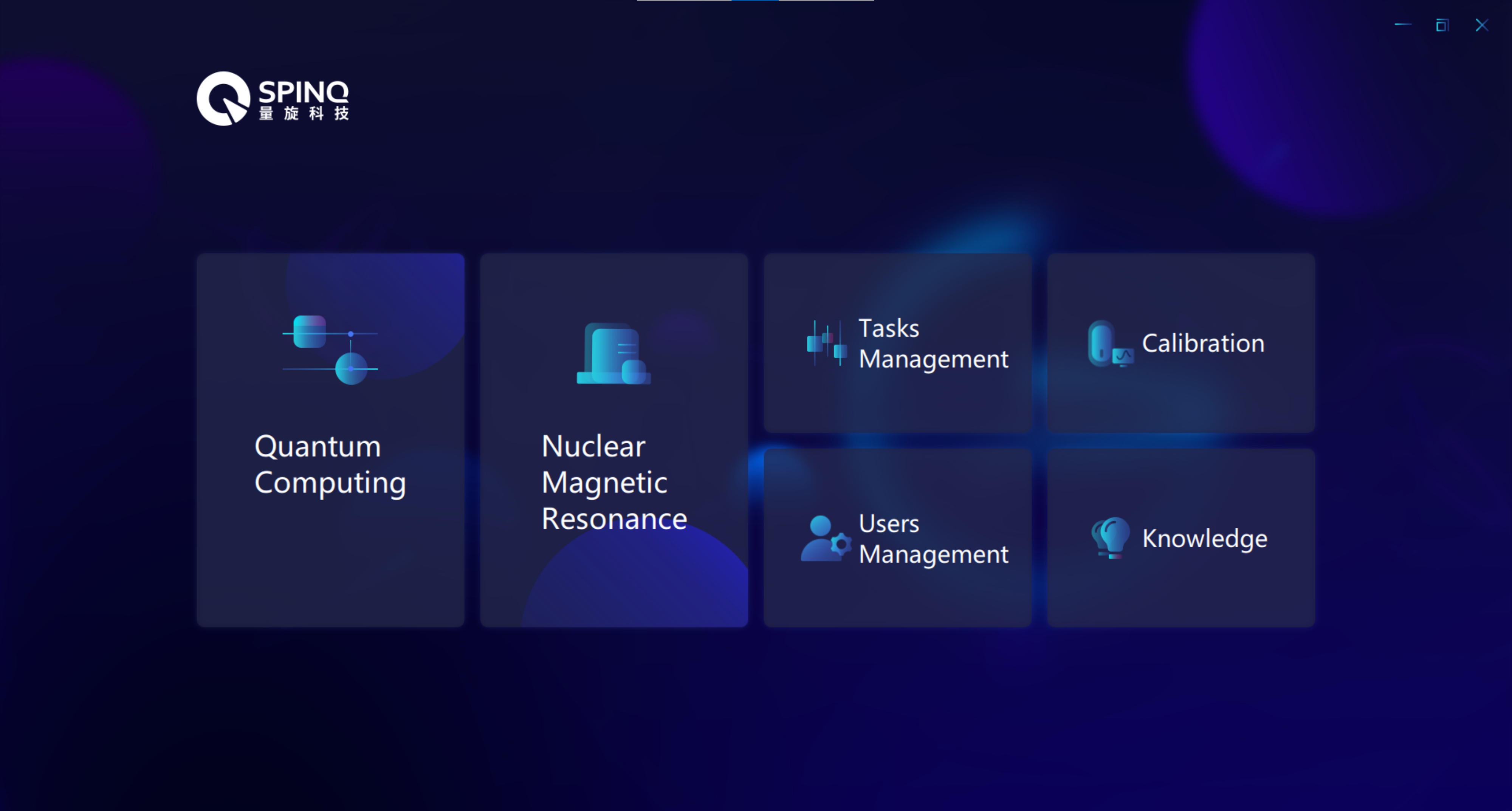}
\medskip
\caption{The homepage of SpinQuasar, the user-interface software for SpinQ Gemini, installed on a personal computer (PC).}\label{figquasar}
\end{figure}

The SpinQ Gemini device comes with the user-interface software SpinQuasar (see Fig. \ref{figquasar}), together forming an integrated hardware-software platform for quantum computing education and research. For further technical details, including how the qubits are physically realized, the reader is referred to \cite{hou}.

We're going to demonstrate the following example on SpinQ Gemini, with the help of SpinQuasar:

\begin{example}\label{exgemini}
Charlie secretly applies a Pauli-$X$ gate on Alice's qubit of an EPR pair $\rho_{AB}=\frac{1}{2}(\vert 00\rangle+\vert 11\rangle)(\langle 00\vert+\langle 11\vert)$. Then, the overall 2-qubit state for Alice and Bob changes to $\rho'_{AB}=\frac{1}{2}(\vert 10\rangle+\vert 01\rangle)(\langle 10\vert+\langle 01\vert)$. However, the state of Alice's individual qubit stays the same: $\rho'_A=\rho_A=\frac{1}{2}\vert 0\rangle\langle 0\vert+\frac{1}{2}\vert 1\rangle\langle 1\vert$. So Alice has no chance to figure out by herself that Charlie did something. But \textit{together} with Bob, they may figure it out!
\end{example}

\begin{figure*}[p]
\centering
\includegraphics[width=0.98\linewidth]{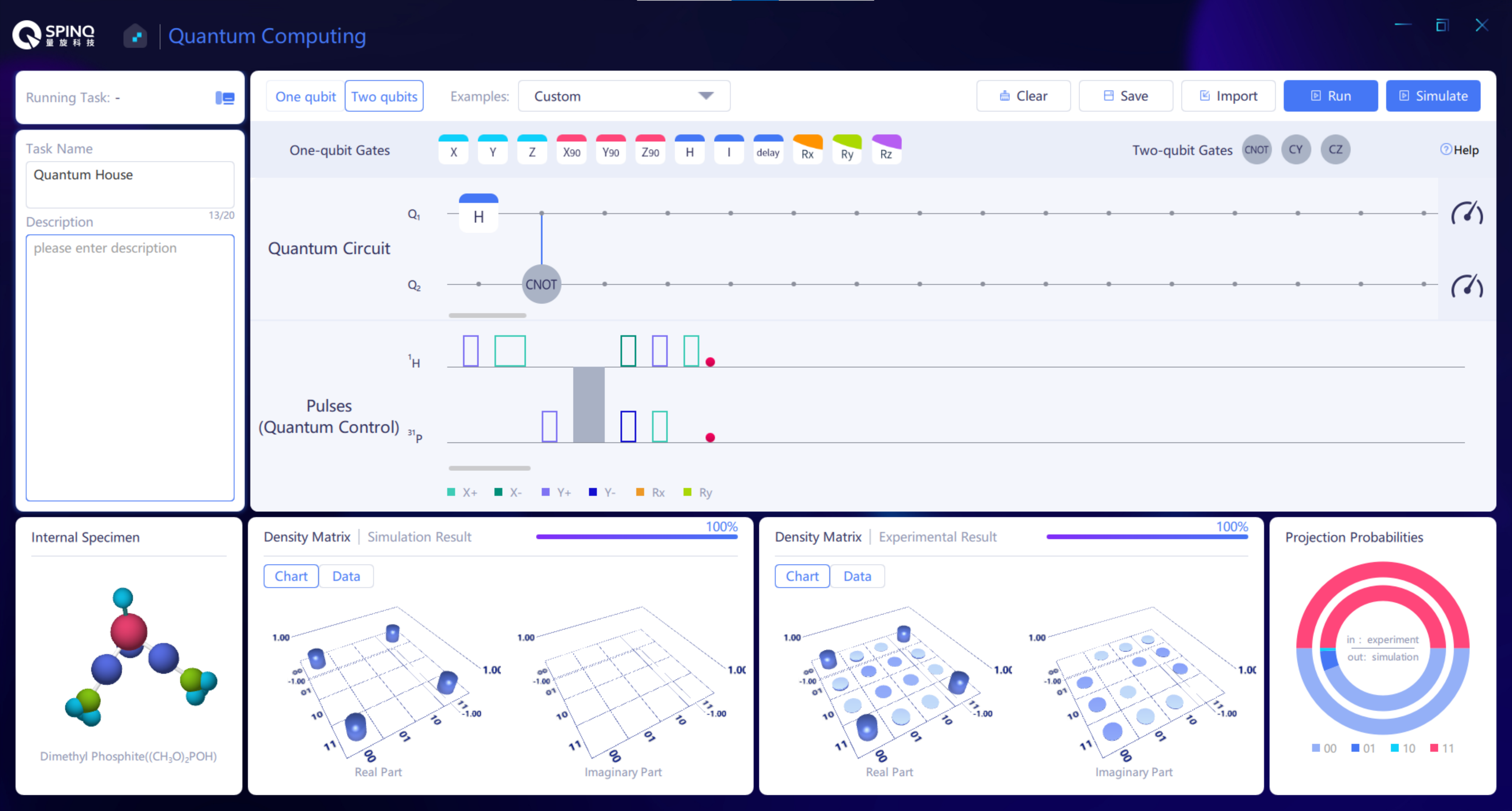}\\
\bigskip
\includegraphics[width=0.98\linewidth]{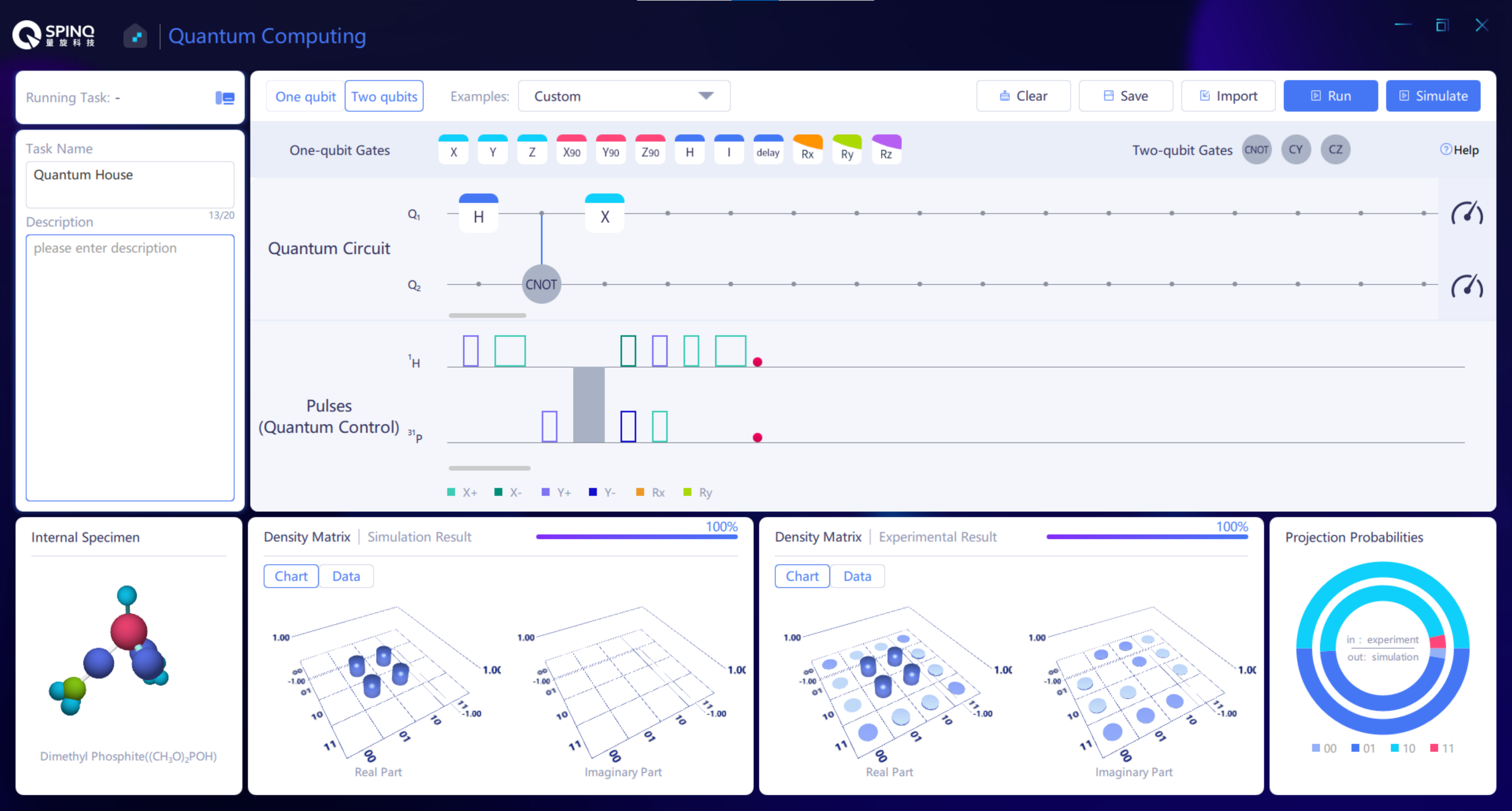}\\
\medskip
\caption{Overall impact of applying a Pauli-$X$ gate on the first qubit of the EPR pair $\rho_{AB}=\frac{1}{2}(\vert 00\rangle+\vert 11\rangle)(\langle 00\vert+\langle 11\vert)$. \textbf{Top screenshot:} preparation of the EPR pair $\rho_{AB}$ in SpinQ Gemini. In the "Quantum Circuit" section, there is a Hadamard gate ($H$) on the first qubit, followed by a CNOT gate; this circuit is executed to produce $\rho_{AB}$. The "Density Matrix $\vert$ Simulation Result" section shows the ideal $\rho_{AB}$ in chart format (see Fig. \ref{figepr2qdata} for numerical format), where four matrix elements (in the four corners) have the value $0.5$, the rest are all zeros. Next to it, the "Density Matrix $\vert$ Experimental Result" section shows the noisy EPR-pair density matrix that is actually produced by the SpinQ Gemini hardware. In spite of the noise, the matrix elements in the four corners are still close to $0.5$, and the rest are all close to zero as well. \textbf{Bottom screenshot:} the 2-qubit state after applying an additional Pauli-$X$ gate on the first qubit of the EPR pair. In the density-matrix charts we can see a clear difference: the $0.5$ values have moved from the corners to the middle. The new ideal state is $\rho'_{AB}=\frac{1}{2}(\vert 10\rangle+\vert 01\rangle)(\langle 10\vert+\langle 01\vert)$, which is different from $\rho_{AB}$.}\label{figepr2q}
\end{figure*}

The SpinQuasar screenshots in Fig. \ref{figepr2q} show how we implemented the EPR pair on SpinQ Gemini, as well as the EPR pair followed by a Pauli-$X$ gate on the first qubit. In each case, SpinQuasar displays not only the ideal, i.e. noiseless, 2-qubit density matrix, but also the noisy density matrix which was actually produced by the hardware.\footnote{Due to the peculiarities of liquid-state NMR technology, whenever we command SpinQ Gemini to produce a pure $n$-qubit state $\rho=\vert\psi\rangle\langle\psi\vert$, such as the EPR pair, the hardware will instead produce a so-called pseudo-pure state $\sigma=(1-\eta)\frac{I}{2^n}+\eta\vert\psi\rangle\langle\psi\vert$, where $\eta\sim 10^{-5}$ for $n=1,2$. This happens under the hood, and as $\rho$ and $\sigma$ are equivalent in the sense that we can unambiguously calculate one from the other, SpinQuasar only shows us $\rho$ (both ideal and noisy), but not $\sigma$.} We can clearly see that applying a Pauli-$X$ gate on the first qubit changes the overall 2-qubit state.

\begin{figure*}[p]
\centering
\includegraphics[width=0.98\linewidth]{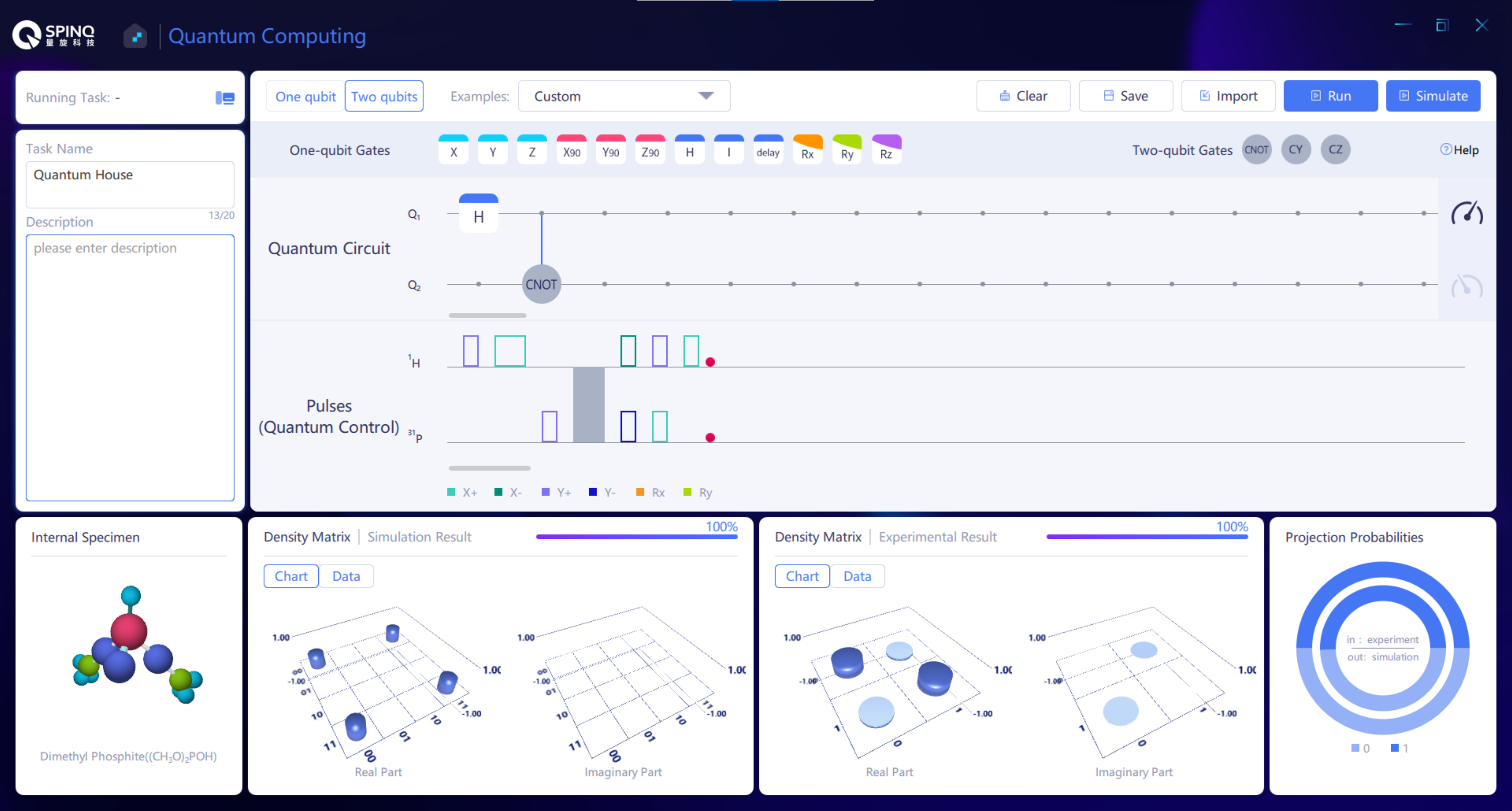}\\
\bigskip
\includegraphics[width=0.98\linewidth]{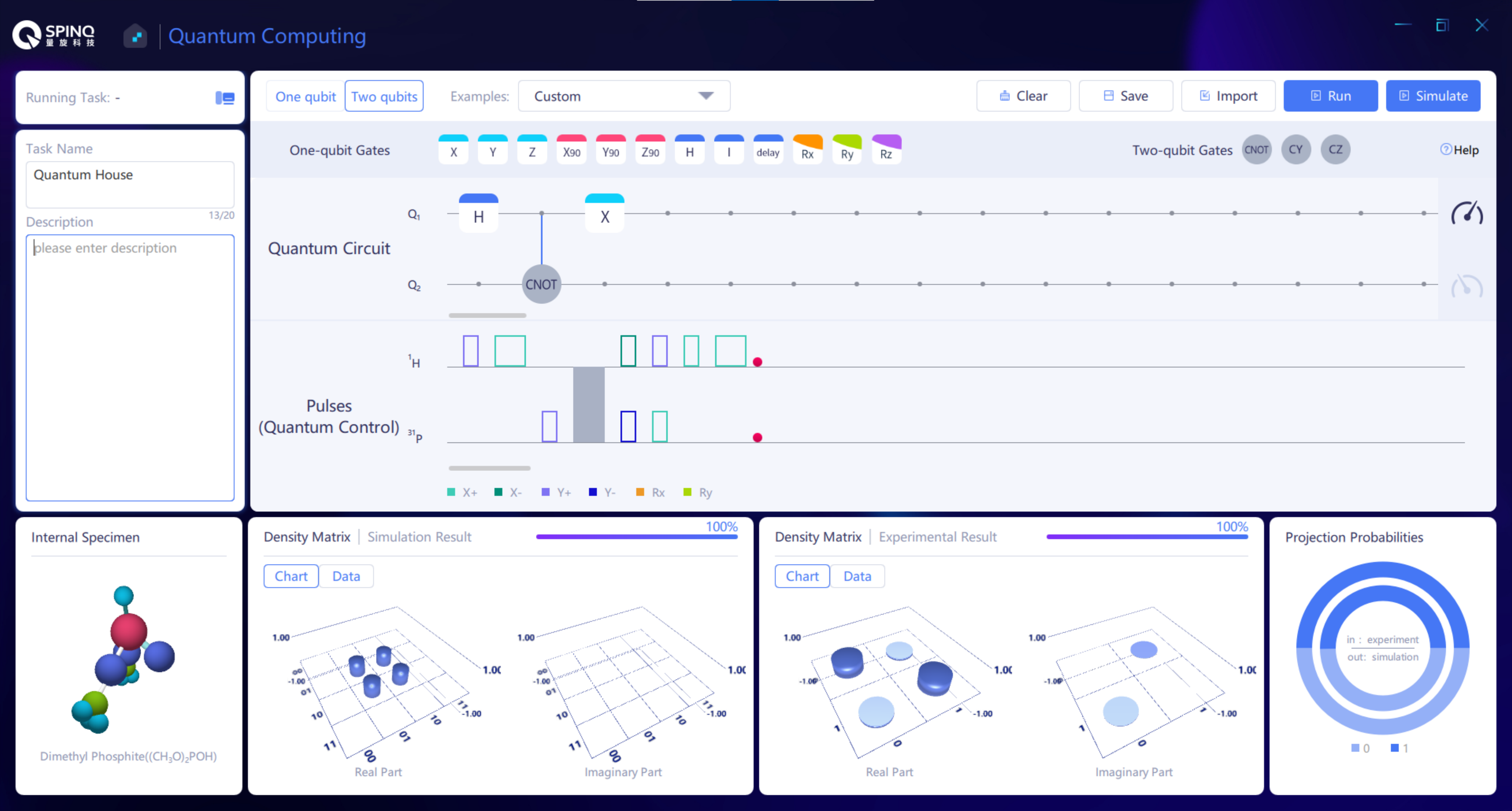}\\
\medskip
\caption{Partial impact of applying a Pauli-$X$ gate on the first qubit of the EPR pair $\rho_{AB}=\frac{1}{2}(\vert 00\rangle+\vert 11\rangle)(\langle 00\vert+\langle 11\vert)$. \textbf{Top screenshot:} the only difference to Fig. \ref{figepr2q} is that here the "Density Matrix $\vert$ Experimental Result" section shows the noisy density matrix chart of the first qubit only (see Fig. \ref{figepr1qdata} for numerical format). It is because the measurement icon at the end of "Quantum Circuit" is greyed out for the second qubit. Similarly to Fig. \ref{figepr2q}, the noisy density matrix is close to the ideal $\rho_A=\mathrm{tr}_B( \rho_{AB})=\frac{I}{2}$, which is the 1-qubit maximally mixed state. On the other hand, the "Density Matrix $\vert$ Simulation Result" section still shows the ideal 2-qubit state $\rho_{AB}$, just like in Fig. \ref{figepr2q}. \textbf{Bottom screenshot:} after applying an additional Pauli-$X$ gate on the first qubit of the EPR pair, the density matrix chart indicates that the state of the first qubit remains the same (apart from noise), namely $\rho'_A=\mathrm{tr}_B( \rho'_{AB})=\frac{I}{2}$, where $\rho'_{AB}=\frac{1}{2}(\vert 10\rangle+\vert 01\rangle)(\langle 10\vert+\langle 01\vert)$.}\label{figepr1q}
\end{figure*}

Then, the screenshots in Fig. \ref{figepr1q} show that as opposed to the overall 2-qubit state, the state of the first qubit alone isn't changed (apart from noise) by applying a Pauli-$X$ gate! And this completes the demonstration of the quantum-house effect.

But there is something we shouldn't overlook here. In fact, due to the noise it would be more accurate to say that we only \textit{illustrated} the quantum-house effect, rather than implemented it on the hardware level. This is because, from a cryptographic point of view, the noise characteristics of the quantum device might give Alice enough hints to be able to figure out whether or not Charlie has applied a Pauli-$X$ gate on the first qubit. So the noise always has to be taken into consideration in a realistic situation.

Therefore, we propose an informal definition for the non-ideal case where noise is present, but won't pursue it further in this paper.

\begin{definition}[Noisy quantum-house effect]\label{defnqhe}
The noisy quantum-house effect is the phenomenon when an operation on subsystem $A$ changes the overall state of system $AB$ significantly, while causing only insignificant change to the state of $A$.
\end{definition}

It's like a non-local, immediate butterfly effect, with the twist that in the extreme, noiseless case, even \textit{no change} to subsystem $A$ causes a significant change to system $AB$.

\section{Theory - Part 2}\label{sec4}

Next, we give a characterization of the $\rho_{AB}$ states with which the quantum-house effect can be achieved. Surprisingly, we'll find that the quantum-house effect is possible even for some product states $\rho_{AB}=\rho_A\otimes\rho_B$, provided that neither $\rho_A$ nor $\rho_B$ are pure states. Loosely speaking, the quantum-house effect is possible whenever there is a non-trivial correlation initially present between $A$ and $B$.\footnote{The ensemble of a pure density $\vert\psi\rangle\langle\psi\vert$ consists only of $\vert\psi\rangle$, so there is exactly one way it can be related to any other ensemble. Put differently, non-trivial correlation is only possible between sets that have multiple elements.}

\begin{theorem}[Non-product implies quantum-house]\label{thmnonprod}
The quantum-house effect can be achieved with any non-product state $\rho_{AB}\ne\rho_A\otimes\rho_B$.
\end{theorem}

\begin{proof}
If we swap $A$ with an independently prepared quantum system in state $\rho_A$, then the new overall state for Alice and Bob will be $\rho'_{AB}=\rho_A\otimes\rho_B$, which is a product state, so clearly $\rho'_{AB}\ne\rho_{AB}$.\footnote{Local operations on $A$ never change the quantum state of $B$, due to the no-signaling principle.} From this, we can also see that the state of Alice's subsystem remained $\rho_A$, and thus we have achieved the quantum-house effect.
\end{proof}

\begin{example}\label{exnonprodqhe}
Let $\rho_{AB}=\frac{1}{2}\vert 00\rangle\langle 00\vert+\frac{1}{2}\vert 11\rangle\langle 11\vert$. This is a non-product state with zero discord (see Example \ref{exdis1}), and a straightforward calculation reveals that $\rho_A=\frac{1}{2}\vert 0\rangle\langle 0\vert+\frac{1}{2}\vert 1\rangle\langle 1\vert=\frac{I}{2}$, the 1-qubit maximally mixed state. Now, if Charlie replaces Alice's qubit with an independently prepared qubit in state $\frac{I}{2}$, then the resulting new overall 2-qubit state for Alice and Bob will be $\rho'_{AB}=\frac{1}{2}\frac{I}{2}\otimes\vert 0\rangle\langle 0\vert+\frac{1}{2}\frac{I}{2}\otimes\vert 1\rangle\langle 1\vert=\frac{I}{2}\otimes\frac{I}{2}$, which is a product state and thus different from $\rho_{AB}$. At the same time, the state of Alice's qubit stays $\rho'_A=\rho_A=\frac{I}{2}$.
\end{example}

\begin{theorem}[Non-trivial correlation implies quantum-house]\label{thmnontrivialqhe}
The quantum-house effect can be achieved with any product state $\rho_{AB}=\rho_A\otimes\rho_B$ where neither $\rho_A$ nor $\rho_B$ is pure.
\end{theorem}

\begin{proof}
Let $\sigma_{A'B'}$ be  a non-product state of some bipartite system $A'B'$ with $\sigma_{A'}=\rho_A$ and $\sigma_{B'}=\rho_B$. Such a state can always be produced, because both $\rho_A$ and $\rho_B$ have support containing more than one element, and thus they can be made classically correlated with each other. We give $B'$ to Bob (i.e. $B=B'$), and keep $A'$ for ourselves. Additionally, we independently prepare another system $A$ in state $\rho_A$, and give that to Alice. Now, the overall state of the system possessed by Alice and Bob is $\rho_{AB}=\rho_A\otimes\rho_B$. Then, if we swap Alice's subsystem $A$ with the $A'$ we kept before, the overall state for Alice and Bob changes to $\rho'_{AB}=\sigma_{A'B'}$, which is different from $\rho_{AB}$. But since the state of Alice's subsystem remains $\rho'_A=\sigma_{A'}=\rho_A$, we have achieved the quantum-house effect.
\end{proof}

An important difference between Theorem~\ref{thmnonprod} and Theorem~\ref{thmnontrivialqhe} is that the proof of the latter requires that there is side-information available which is correlated with Bob's subsystem,\footnote{Thus, the swapping operation in the proof of Theorem~\ref{thmnontrivialqhe} is "local" only in a \textit{geographical} sense, in that it is performed inside Alice's lab.} while for the former theorem it's enough if we just know $\rho_{A}$.

\begin{example}\label{exprodqhe}
Let Charlie first prepare $\sigma_{A'B'}=\frac{1}{2}\vert 00\rangle\langle 00\vert+\frac{1}{2}\vert 11\rangle\langle 11\vert$, which is a non-product state with $\sigma_{A'}=\sigma_{B'}=\frac{I}{2}$. Charlie gives the second qubit to Bob, and keeps the first for himself. Then, he prepares a new qubit, independently in state $\frac{I}{2}$, and gives that to Alice. Thus, for Alice and Bob the overall 2-qubit state is $\rho_{AB}=\frac{I}{2}\otimes\frac{I}{2}$. Now, if Charlie swaps Alice's qubit with the one he kept before, the overall 2-qubit state for Alice and Bob will change to $\rho'_{AB}=\sigma_{A'B'}$, which is different from $\rho_{AB}$. However, the state of Alice's qubit remains the same: $\rho'_A=\sigma_{A'}=\frac{I}{2}$.
\end{example}

Finally, it's easy to see that with the previous theorem we've reached the limit:

\begin{theorem}[Trivial correlation implies no quantum-house]\label{thmtrivialnoqhe}
The quantum-house effect cannot be achieved with any product state $\rho_{AB}=\rho_A\otimes\rho_B$ where either $\rho_A$ or $\rho_B$ is pure.
\end{theorem}

\section{The quantum-house game}\label{sec5}

We present a protocol called "the quantum-house game". The game is played by Alice, Bob and Charlie. Alice has to find out if Charlie (secretly) did something to her subsystem $A$. If she succeeds, she scores points, and her goal is to maximize her expected score.

We'll show that in certain non-classical setups of the game, due to the quantum-house effect Alice is expected to score more points if she asks Bob to help her, while in a classical setup it \textit{never} improves Alice's expected score if she joins forces with Bob.

\subsection{The protocol}\label{subsec51}

The game is rather open-ended, leaving Charlie a lot of freedom in Steps 1, 2 and 3 to shape what "flavor" to play.

\begin{algorithm}[H]
\floatname{algorithm}{Protocol}
\caption{The quantum-house game}\label{protocol1}
\begin{algorithmic}[1]
\State Charlie prepares a bipartite physical system $AB$, and gives subsystem $A$ to Alice and $B$ to Bob. Charlie also provides Alice with information about $AB$. It's up to Charlie what information he shares, if any.
\State Charlie comes to Alice's lab and checks if Alice has tampered with $A$. If she is caught, the protocol is aborted and she scores \textbf{negative infinity points}.
\State Still in Alice's lab, Charlie may or may not, with 50-50\% chance each, secretly perform a pre-agreed operation on subsystem $A$. The operation has to be such that it changes the state of $AB$ when performed.
\State Alice, back to her lab, examines $A$ to figure out whether or not Charlie has performed the operation on it. Afterwards, she either goes to Step 5 to ask Bob for help, or ends the game by taking a guess. In the latter case, she scores \textbf{100 points} if her guess is correct, but \textbf{0 points} if it's incorrect.
\State Alice and Bob examine together the whole system $AB$, after which Alice guesses whether or not Charlie has performed the operation on $A$. She scores \textbf{90 points} if her guess is correct, but \textbf{0 points} if it's incorrect.
\end{algorithmic}
\end{algorithm}

Before going further, we'd like to underline that having a score of "negative infinity" is perfectly fine, as long as we can calculate with it in a mathematically rigorous way. (The score is a logical entity, not physical.) For example, we can use hyperreal numbers \cite{keisler}, an extension of the real numbers which contains infinite numbers as well as infinitesimals.\footnote{On the other hand, we \textit{can} assume that Alice isn't able to physically perform e.g. a rotation gate with an infinitesimal angle.}

\subsection{Classical $AB$}\label{subsec52}

The first thing we notice is that whenever $AB$ is a classical system, Step 5 isn't needed.

It's because Alice can, in principle, observe a classical $A$ without perturbing either $A$ or $AB$. So she won't be caught in Step 2. Thus, she is able to observe the exact state of $A$ both before and after Charlie's action of Step 3.

Now, if Charlie causes in Step 3 a noticeable change to $AB$, then it will also be noticeable by examining solely $A$. So Alice won't need Bob's help (Step 5), as it wouldn't improve her expected score.

\subsection{Non-classical $AB$}\label{subsec53}

As for non-classical $AB$, we analyze a concrete example (i.e. "flavor") of the game.

\textbf{Step 1.} Charlie prepares for Alice and Bob a 2-qubit system $AB$ in the following state:

\begin{multline}\label{eqab1}
\rho_{AB}=\frac{1}{3}\vert 00\rangle\langle 00\vert+\frac{1}{6}\vert 01\rangle\langle 01\vert\\
+\frac{1}{6}\vert 10\rangle\langle 10\vert+\frac{1}{3}\vert 11\rangle\langle 11\vert
\end{multline}

Due to Theorem \ref{thmdis}, $\rho_{AB}$ has zero discord.\footnote{Since it can be written as $\rho_{AB}=\frac{1}{2}\vert 0\rangle\langle 0\vert\otimes(\frac{2}{3}\vert 0\rangle\langle 0\vert+\frac{1}{3}\vert 1\rangle\langle 1\vert)+\frac{1}{2}\vert 1\rangle\langle 1\vert\otimes(\frac{1}{3}\vert 0\rangle\langle 0\vert+\frac{2}{3}\vert 1\rangle\langle 1\vert)$.} In general, a mixed quantum state can be prepared using different ensembles. In this example, Charlie prepares $AB$ using the ensemble given by:

\begin{multline}\label{eqab2}
\rho_{AB}=\frac{1}{6}\vert 00\rangle\langle 00\vert+\frac{1}{6}\vert 11\rangle\langle 11\vert \\
+\frac{1}{6}\vert {+}0\rangle\langle +0\vert+\frac{1}{6}\vert {-}0\rangle\langle -0\vert \\
+\frac{1}{6}\vert {+i}1\rangle\langle {+}i1\vert+\frac{1}{6}\vert {-i}1\rangle\langle {-i}1\vert
\end{multline}

That is, Charlie prepares with equal probability one of $\vert 00\rangle$, $\vert 11\rangle$, $\vert {+}0\rangle$, $\vert {-}0\rangle$, $\vert {+i}1\rangle$, $\vert {-i}1\rangle$, gives the first qubit to Alice and the second to Bob. Thus, Alice gets a qubit whose state vector $\vert\psi\rangle$ is drawn uniformly randomly from the set $S=\{\vert 0\rangle, \vert 1\rangle, \vert {+}\rangle, \vert {-}\rangle, \vert {+i}\rangle, \vert {-i}\rangle\}$.\footnote{As usual, $\vert\pm\rangle=\frac{1}{\sqrt{2}}(\vert 0\rangle\pm\vert 1\rangle)$, $\vert{\pm i}\rangle=\frac{1}{\sqrt{2}}(\vert 0\rangle\pm i\vert 1\rangle)$.}

Let Charlie inform Alice that he used the ensemble in Equation \ref{eqab2} to produce $\rho_{AB}$. It's important to emphasize here that from Alice's perspective, the density matrix of her qubit isn't $\vert\psi\rangle\langle\psi\vert$, but $\rho_A=\mathrm{tr}_B( \rho_{AB})=\frac{I}{2}$, the 1-qubit maximally mixed state.

\textbf{Step 2.} Charlie does the check by measuring Alice's qubit in a basis which does not to perturb the state of the qubit. E.g. if Charlie gave Alice the $\vert {+}\rangle$ state, he will do the check by measuring it in the Hadamard basis. Then, it's easy to see the following:

\begin{theorem}[]\label{thmnochg}
Unless Alice makes sure $\vert\psi\rangle$ isn't perturbed, she will have a non-zero probability of being caught in Step 2, and thus an expected score of negative infinity.
\end{theorem}

Given that just by random guessing Alice would have an expected score of 50 points, she won't attempt to gain more information about the state of her qubit before Step 2, because that would entail a non-zero chance of perturbing $\vert\psi\rangle$, as the states in $S$ are not pairwise orthogonal, they include conjugate bases \cite{wiesner}. This is a key difference compared to the classical case of Subsection \ref{subsec52}.

\textbf{Step 3.} Let the pre-agreed operation, which Charlie may or may not secretly perform, be a Pauli-$X$ gate on Alice's qubit. Based on Equation \ref{eqab1}, a Pauli-$X$ gate brings $\rho_{AB}$ to:

\begin{multline}\label{eqabnew}
\rho'_{AB}=\frac{1}{3}\vert 10\rangle\langle 10\vert+\frac{1}{6}\vert 11\rangle\langle 11\vert\\
+\frac{1}{6}\vert 00\rangle\langle 00\vert+\frac{1}{3}\vert 01\rangle\langle 01\vert
\end{multline}

Clearly, $\rho_{AB}\ne\rho'_{AB}$. However, it doesn't change the state of Alice's qubit: $\rho'_A=\mathrm{tr}_B( \rho'_{AB})=\frac{I}{2}$, same as $\rho_A$.

\textbf{Step 4.} As we've just seen,  even if Charlie decides to secretly apply a Pauli-$X$ gate, it only changes the state of $AB$, but not that of $A$. Thus, if Alice ends the game here, all she can do is random guessing, resulting in an expected score of 50 points. But we'll see she can do better if she joins forces with Bob. So she goes to Step 5.

\textbf{Step 5.} Alice and Bob measure $AB$ in the computational basis. If the result is $00$ or $11$, Alice guesses that Charlie did nothing. On the other hand, if the result is $01$ or $10$, Alice guesses that Charlie has applied the Pauli-$X$ gate. This way, Alice will be correct with probability $\frac{2}{3}$, giving an expected score of 60 points. So it's worth asking Bob for help.

\subsection{Discussion: implicit discord}\label{subsec54}

We saw in Subsection \ref{subsec53} that in a quantum setting, Alice could make use of the quantum-house effect to achieve a higher expected score by joining forces with Bob. We also explained in Subsection \ref{subsec52} why the same can never happen in a classical setting. Furthermore, in the example we analyzed, $\rho_{AB}$ had zero discord. So, apparently, non-classical correlation was possible without quantum discord. But was it really the case? We'll argue that the \textit{ability} of Charlie's device to create quantum discord was in fact necessary.

To start with, imagine Alice knows in the quantum-house game that Charlie's device isn't able to produce non-vanishing discord, i.e. any $\rho_{AB}$ such that $\delta_{AB}>0$. Based on Subsection \ref{subsec21}, this means that the device cannot create two non-orthogonal states of $A$. Thus, Charlie can only prepare $A$ in itself, in one of the states of a fixed orthonormal basis $\{\vert a_i\rangle\}$. If Alice is aware of that, she can simply measure $A$ in that basis, gaining information without being caught.

Now, in particular, let Alice know in the example of Subsection \ref{subsec53} that Charlie's device is only capable of producing the $\{\vert 0\rangle,\vert 1\rangle\}$ basis states of $A$. So if she measures her qubit in this basis before Step 2, then she can find out already in Step 4 with certainty (by measuring again), whether or not Charlie has applied the Pauli-$X$ gate in Step 3. Thus, without the implicit presence (ability) of quantum discord, the quantum-house effect wouldn't make a difference here, because Alice would never need Step 5.

In general, if in a physical experiment it's only possible to prepare $A$ in itself, in one of the pairwise orthogonal states $\{\vert a_i\rangle\}$, then for all intents and purposes $A$ can be considered as classical. In this sense, we can say that (the possibility of) non-vanishing quantum discord is necessary for quantumness. Or, based on Theorem \ref{thmdis}, it's eventually the possibility of superposition that is necessary.

\section{Quantum detachment}\label{sec6}

In this paper, we introduced the quantum-house effect, a non-local phenomenon which can be exhibited even with bipartite product states, provided that neither subsystem is in a pure state. The effect was demonstrated (with some inevitable noise) on the SpinQ Gemini 2-qubit liquid-state NMR desktop quantum computer.

We also argued that although the effect apparently doesn't require quantum discord to be present, the \textit{ability} to create discord is necessary in a cryptographic (as well as physical) sense, so that the quantum-house effect can make a difference compared to classicality.\footnote{In the mathematical sense, if we take the formal definition of quantum state at face value, no implicit quantum discord is needed, that's what we saw in Sections \ref{sec2} and \ref{sec4}.} This isn't actually surprising if we consider that implicit discord is basically the possibility of superposition.

To go beyond the quantum-house effect, we can view quantumness from a slightly different angle. Let's take the BB84 protocol \cite{bennett} as an example. It merely uses single qubits in pure states and achieves something that is classically unattainable: unconditionally secure communication. What makes this possible in BB84 is that when Alice sends a qubit to Bob, she holds back relevant information about the qubit's state (she can do that because her device is capable of creating non-orthogonal qubit states), which renders the eavesdropper's task impossible. We suggest calling this phenomenon \textit{quantum detachment}, a principle which roughly means that relevant information\footnote{Information that would influence what outcomes an experimenter may expect when physically interacting with the system.} about the state of a physical system is kept separate from the system itself.

In our opinion, one way to characterize the point where quantumness departs from classicality is the presence of quantum detachment. The idea can be conveyed as follows: when the (locally unavailable) information is somewhere else, we have a mixed state; and when it's nowhere else, we have superposition. The latter case can be considered as the ultimate quantum detachment, because the missing information doesn't even exist in the universe.

We can contrast quantum detachment with the classical world where no relevant information about the state of a physical system can be held back, as all of it can be found out locally in the lab, in principle. Put differently, in the classical world a physical system contains all the relevant information about itself.

As for future work, the quantum-house game might be turned into a protocol by which Charlie could securely cast a "yes/no" vote, locally in Alice's lab.

\backmatter

\section*{Declarations}

At the time of writing, Papafut Quantum is an exclusive agent of SpinQ in Switzerland.

\begin{figure*}[p]
\centering
\includegraphics[width=0.98\linewidth]{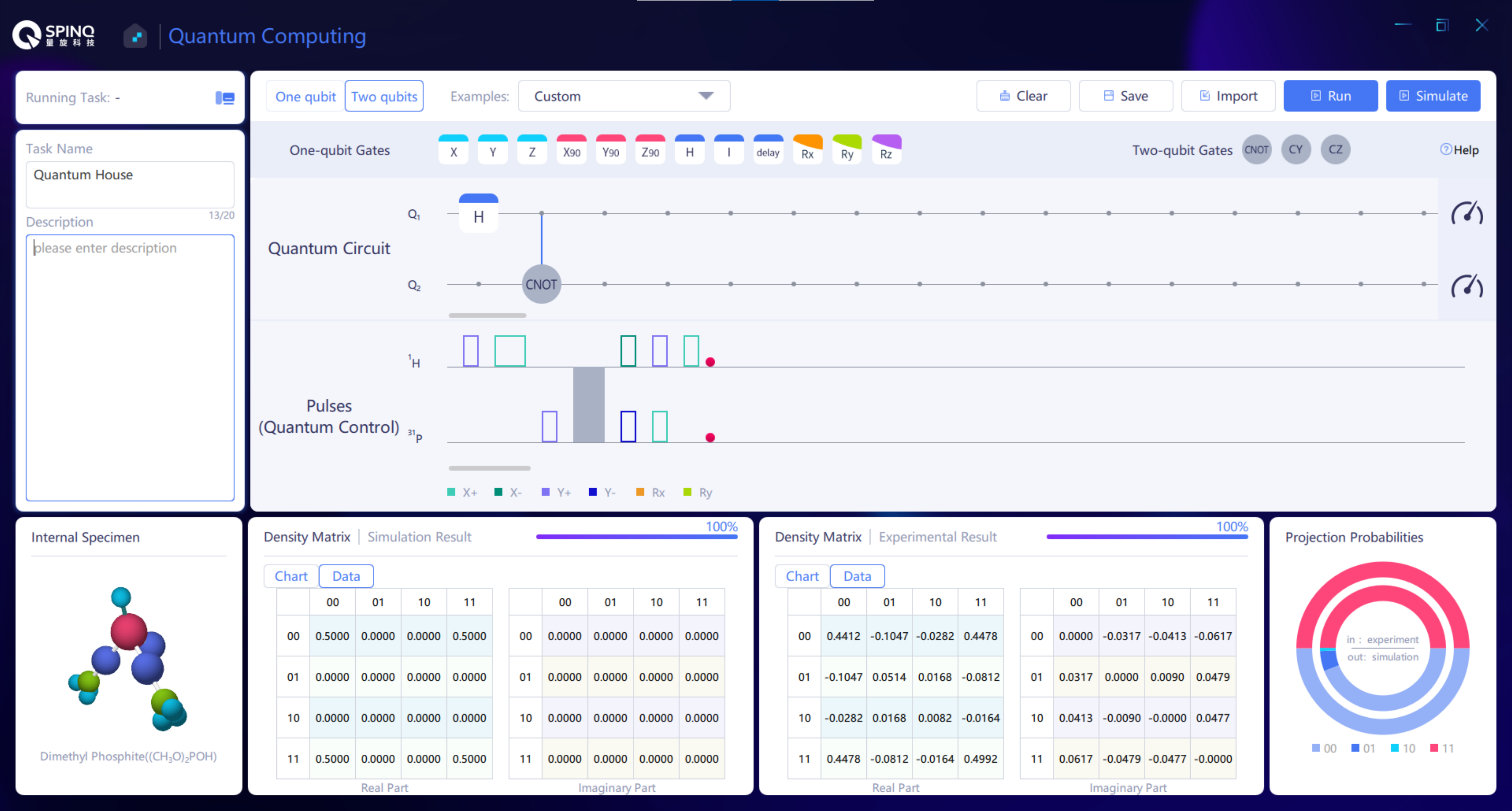}\\
\bigskip
\includegraphics[width=0.98\linewidth]{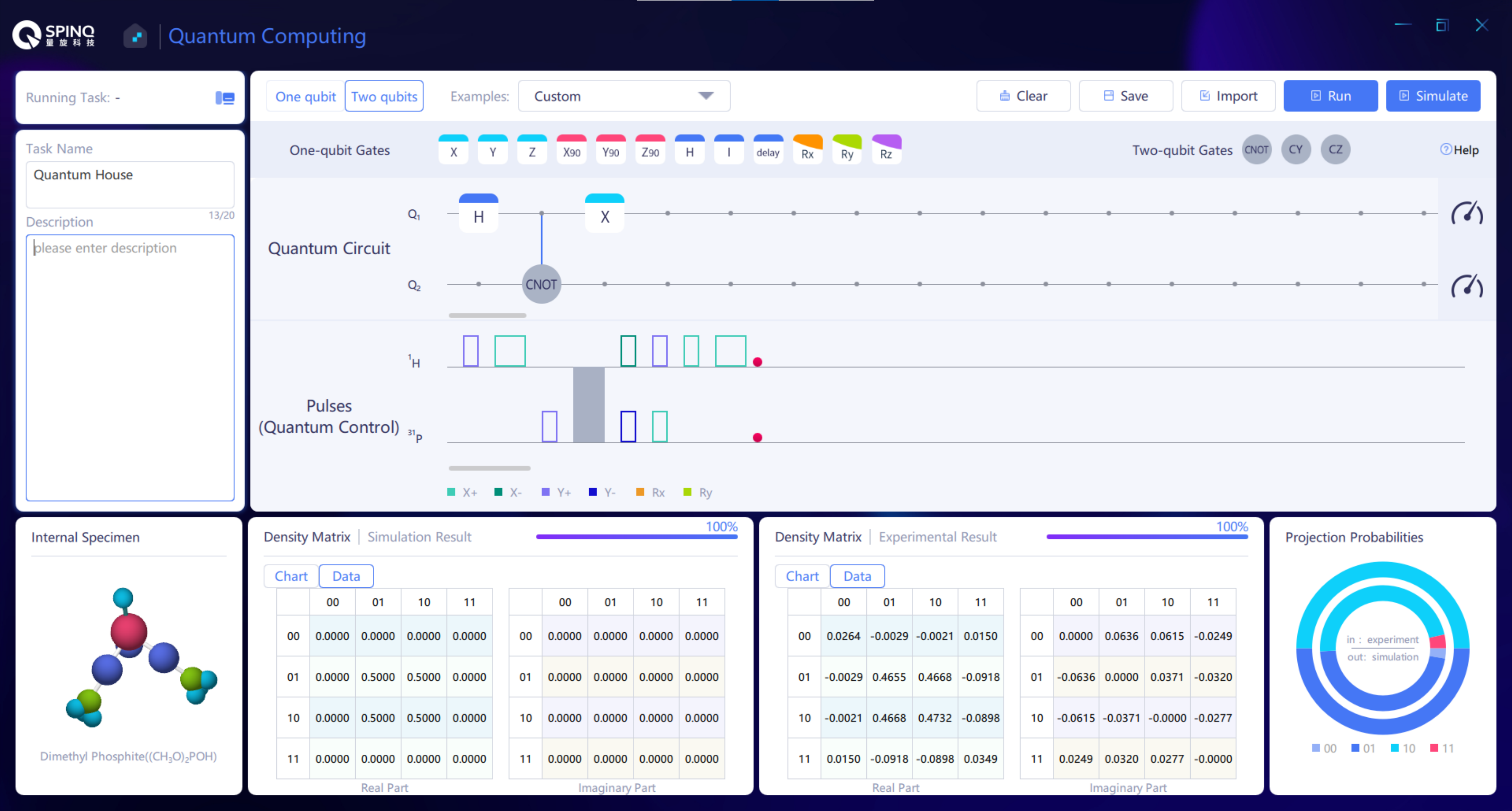}\\
\medskip
\caption{The "Data" view of Fig. \ref{figepr2q}, showing the numerical density matrices.}\label{figepr2qdata}
\end{figure*}

\begin{figure*}[p]
\centering
\includegraphics[width=0.98\linewidth]{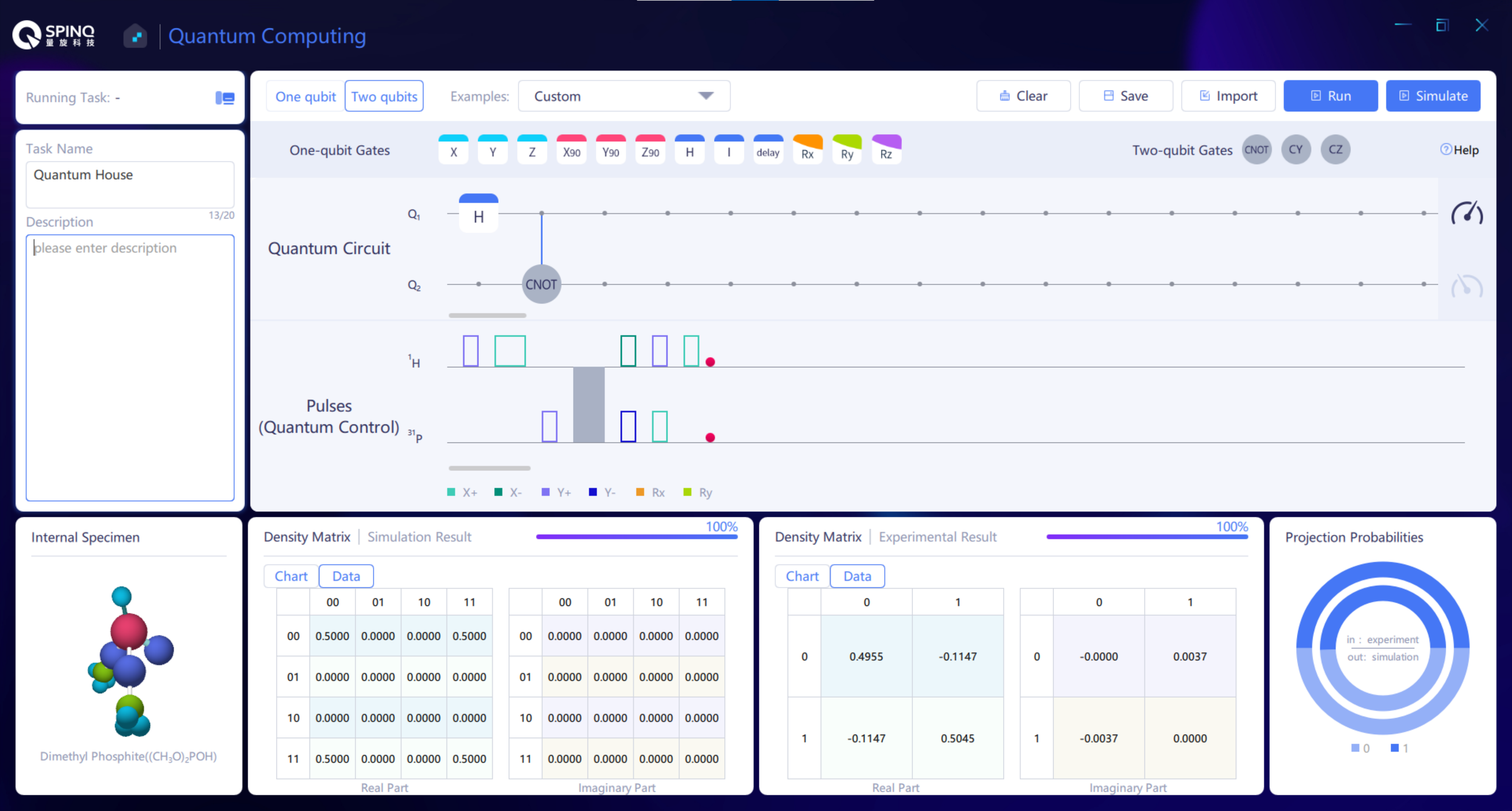}\\
\bigskip
\includegraphics[width=0.98\linewidth]{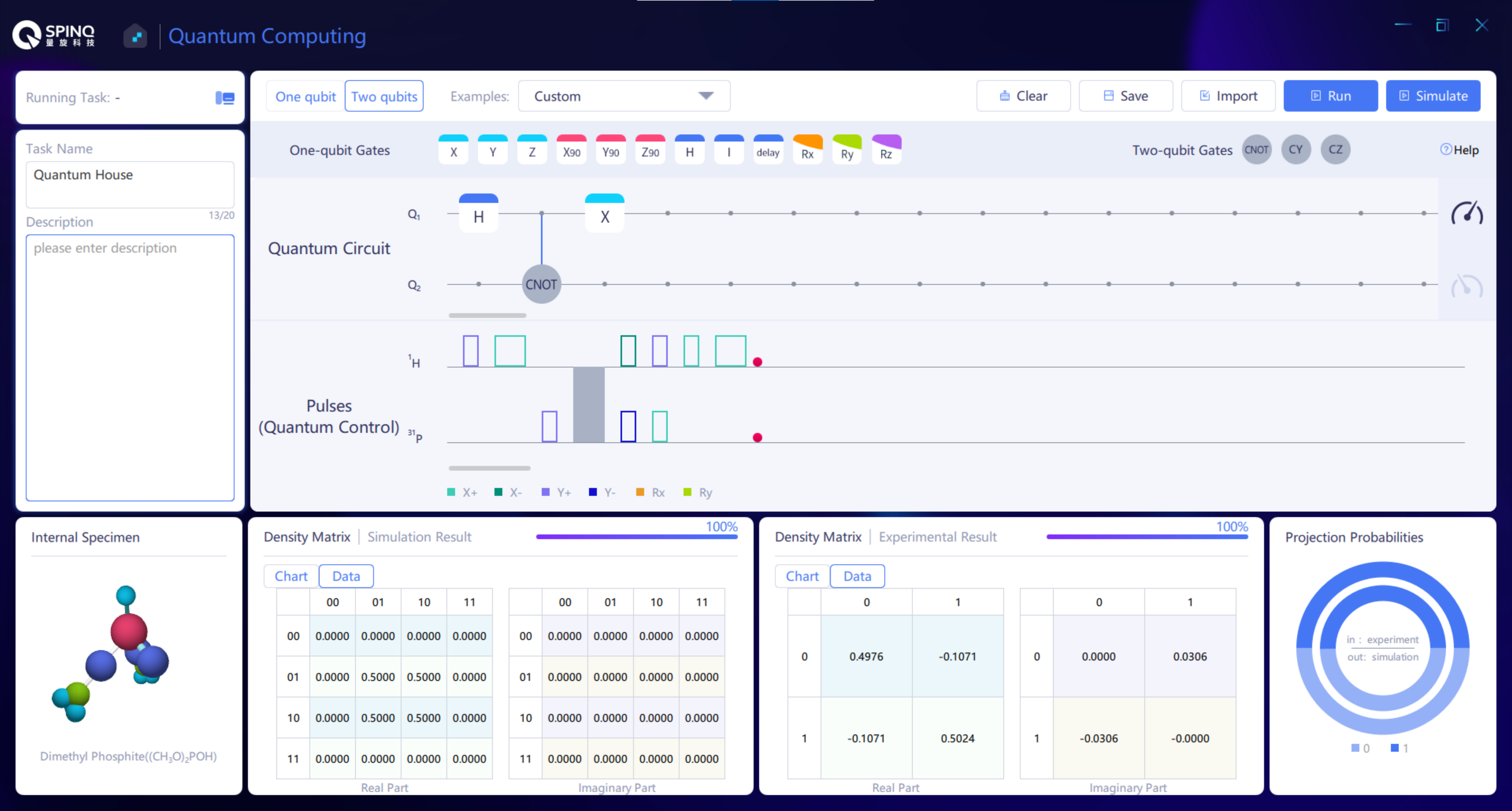}\\
\medskip
\caption{The "Data" view of Fig. \ref{figepr1q}, showing the numerical density matrices.}\label{figepr1qdata}
\end{figure*}

\end{document}